\documentclass[twocolumn,superscriptaddress,floatfix,longbibliography,aps,pra,nopreprintnumbers,10pt]{revtex4-2}

\usepackage[utf8]{inputenc}
\usepackage[english]{babel}
\usepackage{graphicx}
\usepackage{bm}
\usepackage{xcolor}
\usepackage{amsmath}
\usepackage{amssymb}
\usepackage{epstopdf}
\usepackage[normalem]{ulem} 
\usepackage{enumerate}
\usepackage[urlcolor=blue,colorlinks=true,citecolor=blue,linkcolor=blue,pdfstartview={FitH},bookmarks=false]{hyperref}
\renewcommand{\vec}[1]{\mathbf{#1}}
\newcommand{\vA}{\vec{A}}
\newcommand{\vB}{\vec{B}}
\renewcommand{\vr}{\vec{r}}

\newcommand{\vtau}{\mbox{\boldmath $\tau$}}

\begin{document}

\title{Hinge States of Second-order Topological Insulators as a Mach-Zehnder Interferometer}

\author{Adam Yanis Chaou}
\affiliation{Dahlem Center for Complex Quantum Systems and Institut f\"ur Physik,
Freie Universit\"at Berlin, Arnimallee 14, D-14195 Berlin, Germany}

\author{Piet W.\ Brouwer}
\affiliation{Dahlem Center for Complex Quantum Systems and Institut f\"ur Physik,
Freie Universit\"at Berlin, Arnimallee 14, D-14195 Berlin, Germany}

\author{Nicholas Sedlmayr}
\email[e-mail: ]{sedlmayr@umcs.pl}
\affiliation{Institute of Physics, Maria Curie-Sk\l{}odowska University, Plac Marii Sk\l{}odowskiej-Curie 1, PL-20031 Lublin, Poland}

\date{\today}

\begin{abstract}
Three-dimensional higher-order topological insulators can have topologically protected chiral modes propagating on their hinges. Hinges with two co-propagating chiral modes can serve as a ``beam splitter'' between hinges with only a single chiral mode. Here we show how such a crystal, with Ohmic contacts attached to its hinges, can be used to realize a Mach-Zehnder interferometer. We present concrete calculations for a lattice model of a first-order topological insulator in a magnetic field, which, for a suitable choice of parameters, is an extrinsic second-order topological insulator with the required configuration of chiral hinge modes.
\end{abstract}

\maketitle

\section{Introduction}

A Mach-Zehnder interferometer gives an interference pattern which depends on the phase shift between two paths taken by a beam-split signal. Optical Mach-Zehnder interferometers have a long history~\cite{Born2019}. Taking advantage of the absence of backscattering resulting from the unidirectional electron motion in the edge states of the two-dimensional integer quantized Hall effect, a Mach-Zehnder interferometer could also be realized in a two-dimensional electron gas \cite{Ji2003a,Neder2007}. In this case, the electronic equivalent of a ``beam splitter'' is formed by a point contact, which allows for the controllable coupling of modes at different sample edges.

One-dimensional electron modes without backscattering also exist at the hinges of a three-dimensional second-order topological insulator~\cite{Volovik2010,Sitte2012,Zhang2013b,Benalcazar2017a,Langbehn2017,Song2017,Schindler2018,Fang2019,Trifunovic2021,Xie2021}. Signatures of such hinge modes have been seen in pure Bismuth \cite{Schindler2018a,Nayak2019}, in Bi-based compounds \cite{Noguchi2021,Aggarwal2021}, and in Fe-based superconductors~\cite{Gray2019,Zhang2019d}. Interference effects involving pairs of counter-propagating (``helical'') \cite{Niyazov2018} or unidirectional (``chiral'') \cite{Luo2021,Li2021} hinge modes were proposed theoretically. In these proposals, the beam splitter is formed by the point contact between an idealized single-channel normal-metal lead and the topological insulator. In this article, we show how a crystal hinge supporting multiple chiral modes naturally forms a beam splitter between adjacent hinges with only a single mode. This way, a Mach-Zehnder interferometer can be realized with Ohmic source and drain contacts placed over a crystal hinge.

\begin{figure}
  \centering
    \includegraphics[width=0.49\columnwidth]{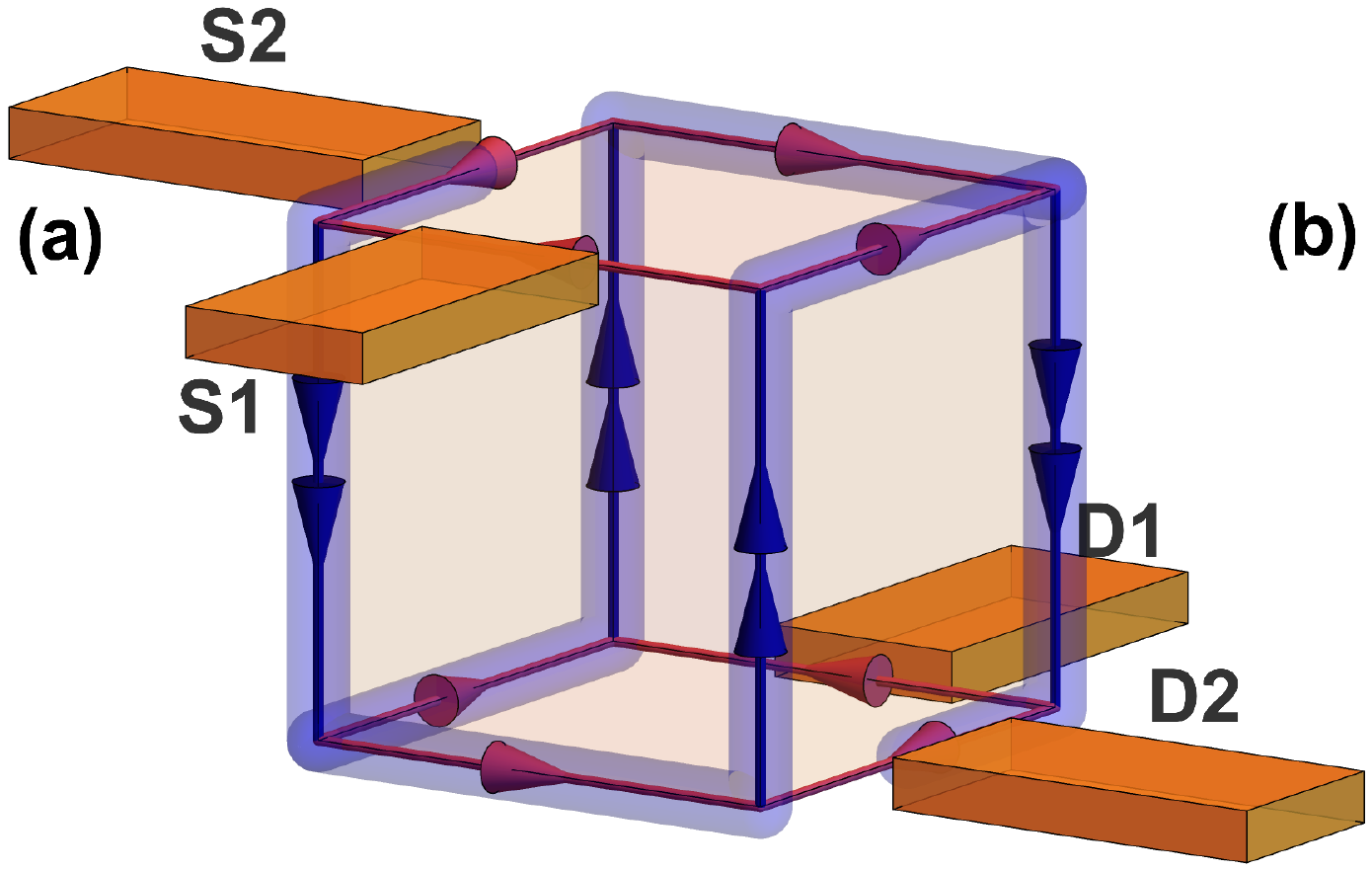}
    \includegraphics[width=0.49\columnwidth]{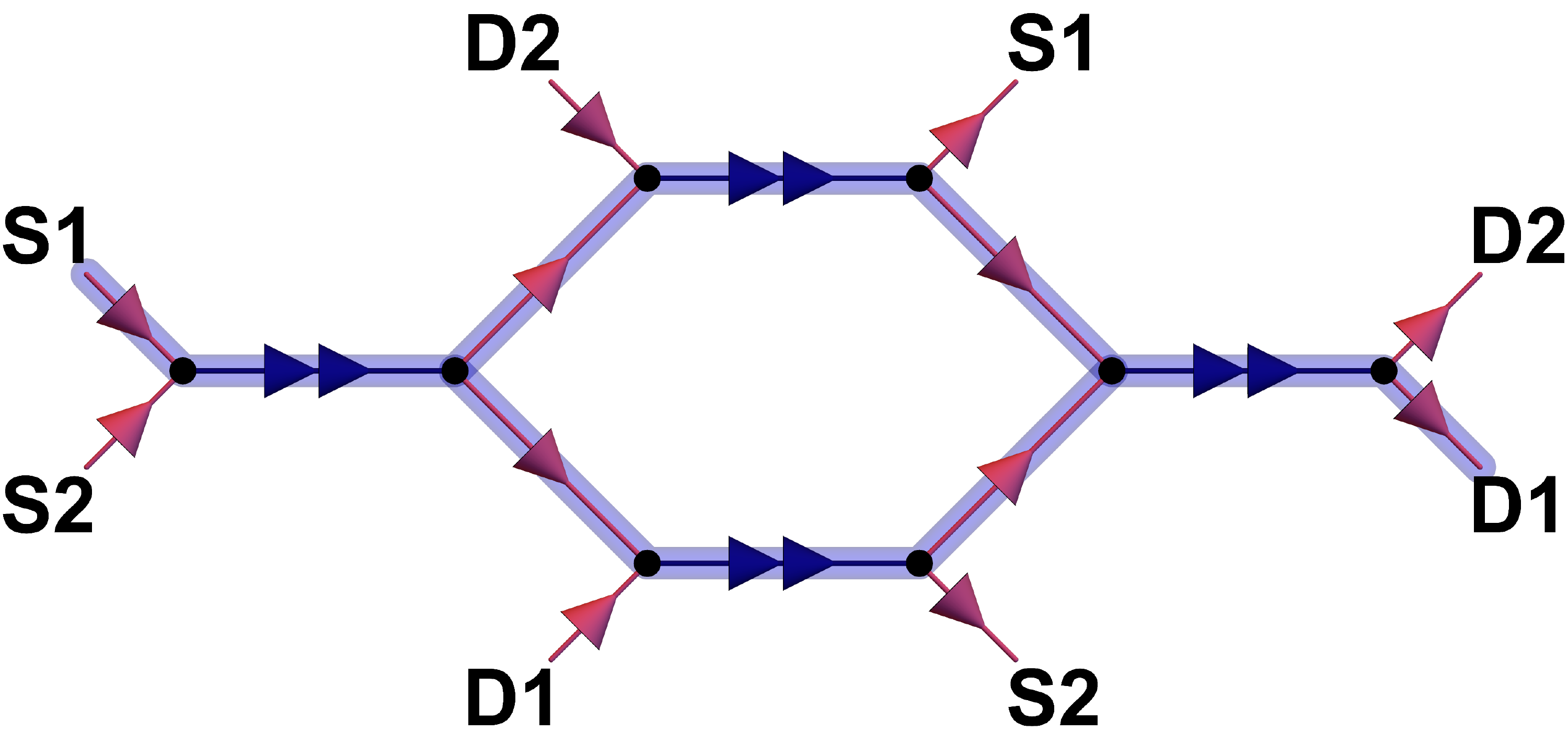}
    \caption{(a) A schematic of the system considered in this article. It consists of a second-order topological insulator with hinges that have one or two chiral states. Ohmic contacts, which pairwise serve as source (S1, S2) and drain (D1, D2) contacts, are attached to four of the hinges with one chiral mode. A pair of interfering paths connecting contacts S1 and D1 is indicated in blue. (b): Effective network diagram indicating the same interfering paths between contacts S1 and D1.}
    \label{fig:schematic}
\end{figure}

The setup we consider is shown schematically in Fig.\ \ref{fig:schematic}. It consists of a second-order topological insulator with hinges that have one or two chiral modes, as indicated in the figure. The crystal hinges with two co-propagating chiral modes serve as beam splitters. Ohmic source and drain contacts are placed at selected crystal edges with a single chiral hinge mode, such that there are two paths connecting each pair of source and drain contacts along the crystal hinges. By controlling the phase difference between the interfering paths with a magnetic field one thereby obtains a Mach-Zehnder interferometer.

Like the chiral edge states of the integer quantized Hall effect, the hinge modes of a higher-order topological insulator are topologically protected. One distinguishes ``intrinsic'' hinge modes, which are protected by the topology of the bulk band structure and ``extrinsic'' modes, for which the nontrivial topology resides in the surface band structure, whereas the bulk may be topologically trivial \cite{Geier2018,Trifunovic2018,Trifunovic2021}. Whereas intrinsic higher-order phases require crystalline symmetries for their protection, extrinsic topological phases do not have additional symmetry requirements. For the realization of an interferometer, all that matters is the existence of the chiral hinge modes, not where they derive their protection from. For that reason, in this article we  seek a (theoretical) realization of a Mach-Zehnder interferometer in an extrinsic second-order topological insulator.

\begin{figure}
    \includegraphics[width=0.995\columnwidth]{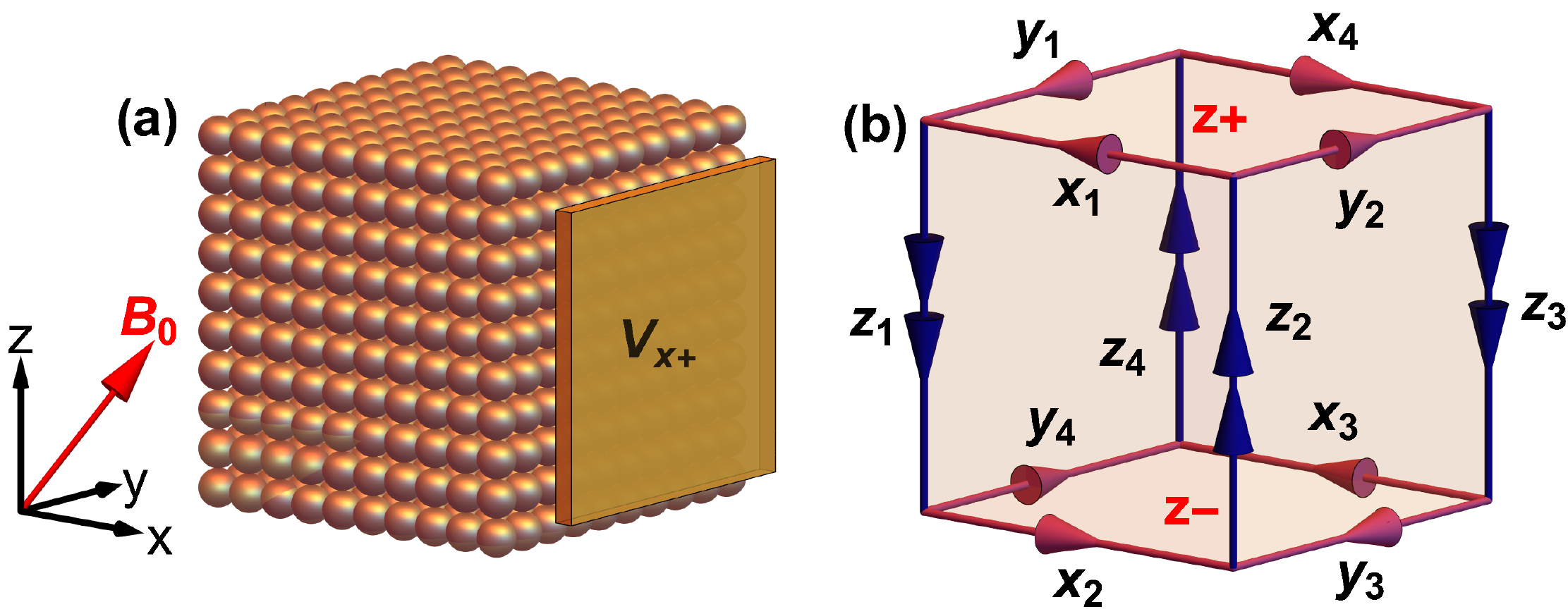}
  \caption{(a): An extrinsic second-order topological insulator with chiral hinge modes can be realized by placing a first order topological topological insulator in a uniform magnetic field $\vB_0$. The direction of $\vB_0$ is chosen such that there is a nonzero flux through all crystal surfaces. As a result, the Dirac-cone surface states form an integer quantized Hall effect with a filling fraction that depends on the magnetic flux density and the electron density at the surface. The latter can be controlled capacitively by a metal gate, one is shown explicitly on surface $x+$. (b): Labeling of surfaces and hinges. The surfaces on the sides are labelled by $x\pm$ and $y\pm$ analogously to $z\pm$, labelled here in red.}
    \label{fig:model}
\end{figure}

A particularly simple and controllable model of an extrinsic second-order topological insulator was proposed by Sitte {\em et al.} in Ref.\ \cite{Sitte2012}. It consists of a (first-order) topological insulator placed in a magnetic field $\vB_0$ at a generic direction with respect to the crystal faces, see Fig.\ \ref{fig:model}. Without the magnetic field, there are Dirac-cone surface states at the crystal surfaces. The magnetic field $\vB_0$ gaps these out. This effectively turns the crystal surfaces into two-dimensional quantized Hall systems with a (half-integer) filling fraction that depends on the perpendicular component of the magnetic field and the position of the Fermi level with respect to the Dirac point of the surface band structure. The former can be controlled by the applied magnetic field, the latter by a gate voltage applied locally at the surface. The number of hinge states then follows as the difference of the filling fractions of the two adjacent surfaces. To bring about the interference pattern, one considers a small change $\delta \vB$ of the magnetic field. If $|\delta \vB| \ll \vB_0$, $\delta \vB$ changes the phases which electrons pick up while propagating along the hinges, while not affecting the number of hinge states and their properties.

The remainder of this article is organized as follows: In Sec.\ \ref{sec:2} we present a simple lattice model of an extrinsic second-order topological insulator as discussed above and establish that it has the phenomenology shown in Fig.\ \ref{fig:schematic} for a suitable choice of parameters. In Sec.\ \ref{sec:3} we add Ohmic contacts to crystal edges with a single chiral mode, as indicated in Fig.\ \ref{fig:schematic}, and theoretically describe the resulting interferometer setup using scattering theory. In Sec.\ \ref{sec:4} we consider a two-terminal Aharonov-Bohm interferometer based on the same model system. We conclude in Sec.\ \ref{sec:5}. Further details and supporting material can be found in the appendices.

\section{Lattice model of an extrinsic second-order topological insulator}
\label{sec:2}

We theoretically describe the extrinsic second-order topological insulator using a four-band lattice model with nearest-neighbor hopping~\cite{Sitte2012}. It has the Hamiltonian
\begin{equation}
  \hat H = \sum_{\langle i, j\rangle} \hat c_i^{\dagger} t(\vr_i,\vr_j) \hat c_j + \sum_{i}
    \hat c_i^{\dagger} u(\vr_i) \hat c_i,
\end{equation}
where the indices $i$ and $j$ run over all neighboring sites of a three-dimensional simple cubic lattice, $\vr_i$ and $\vr_j$ are the corresponding position vectors, $\hat c_{i}$, $\hat c_{j}$ and $\hat c_{i}^{\dagger}$, $\hat c_{j}^{\dagger}$ are four-component spinor annihilation and creation operators, and $t(\vr_i,\vr_j)$ and $u(\vr_i)$ are $4 \times 4$ matrices. We consider a lattice of size $L_x \times L_y \times L_z$, with surfaces perpendicular to the coordinate axes, shown schematically in Fig.\ \ref{fig:model}. For the nearest-neighbor term $t$ we take
\begin{align}
  t(\vr,\vr') =&\, -\frac{t}{2} \left[ \sigma_3 \tau_0 + \frac{i}{a} \sigma_1 \vtau \cdot (\vr - \vr') \right]
  \nonumber \\ &\, \ \ \mbox{} \times
  e^{i e (\vA(\vr) + \vA(\vr')) \cdot (\vr - \vr')/2 \hbar c},
\end{align}
where $a$ is the lattice constant, $t$ a hopping amplitude (with the dimension of energy), $\sigma_{\alpha}$ and $\tau_{\alpha}$, $\alpha = 1,2,3$, are Pauli matrices, and $\vA(\vr)$ is the vector potential corresponding to the uniform applied magnetic field $\vB$. The on-site term $u$ is
\begin{align}
  u(\vr) = (3 + m) t \sigma_3 \tau_0 + V(\vr) \sigma_0 \tau_0,
\end{align}
where $m$ is a parameter governing the bulk band structure and $V(\vr)$ a scalar potential, which is nonzero in the vicinity of the crystal boundaries only.

Without applied magnetic field, the system has time-reversal symmetry. It is in a topological phase with gapless Dirac-cone surface states for $-2 < m < 0$. The surface Dirac nodes are at zero energy if the scalar potential $V$ is zero, but they may be pushed away from zero by application of uniform potential at the surface. We take a scalar potential of the form
\begin{equation}
  V(\vr) = \sum_{s} V_s e^{-r_{s,\perp}/\xi_s},
  \label{eq:VrVs}
\end{equation}
where the summation index $s$ runs over all six surfaces of the crystal, $V_s$ is a gate voltage at surface $s$, $r_{s,\perp}$ is the distance to the surface $s$, and $\xi_s$ a decay length. In our calculations, we set $m = -1$ and $\xi_s = 5 a$ throughout. 

A uniform magnetic field at a direction such that there is a finite flux penetrating all six crystal surfaces, gaps out the surface Dirac cones and effectively turns the surfaces into gapped quantized Hall effects. The filling fractions of the different surfaces $s$ can be tuned by varying the surface gate voltages $V_s$ of Eq.\ \eqref{eq:VrVs}.

\begin{figure}
  \centering
    \includegraphics[width=\columnwidth]{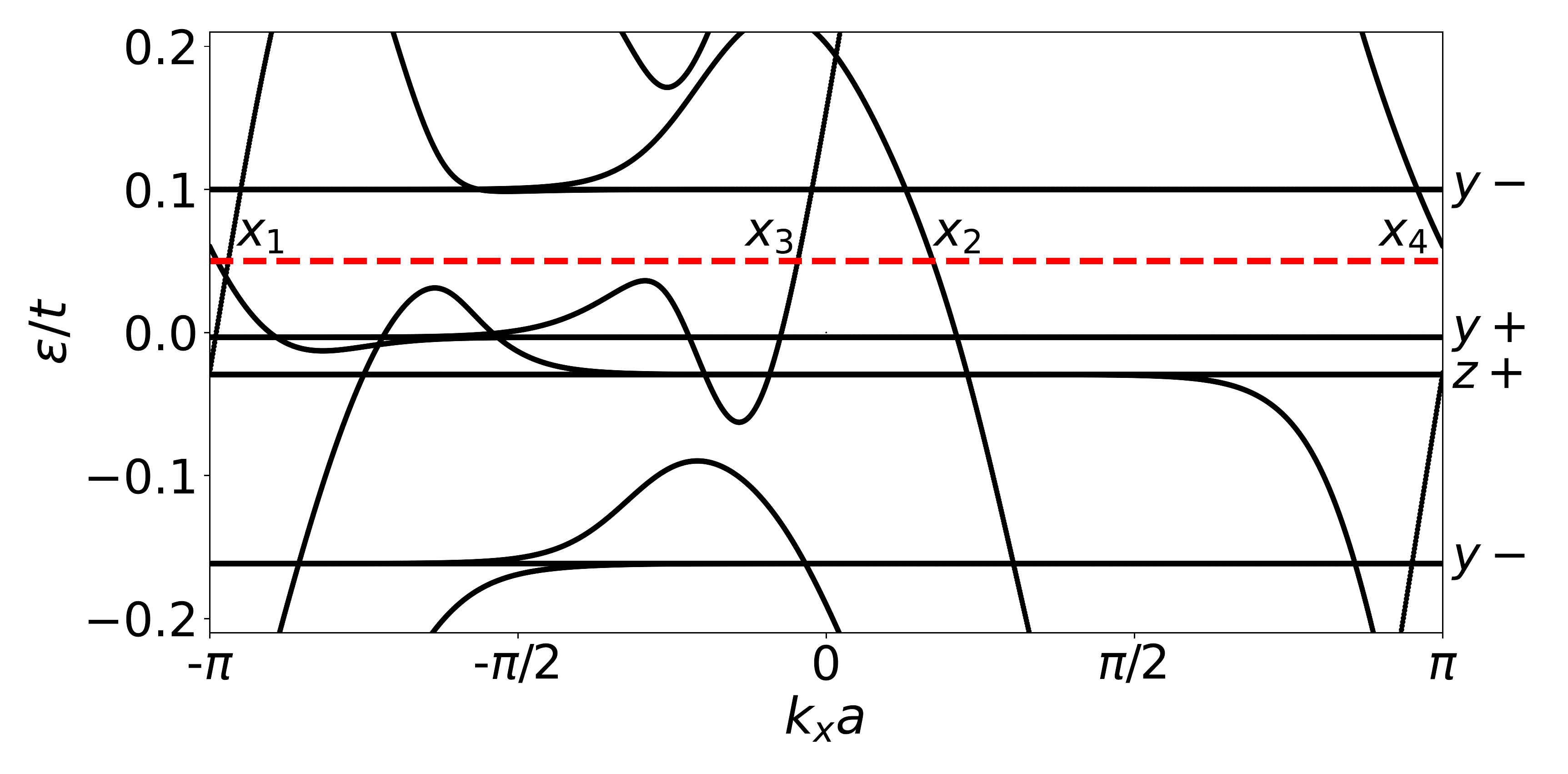}\\
    \includegraphics[width=\columnwidth]{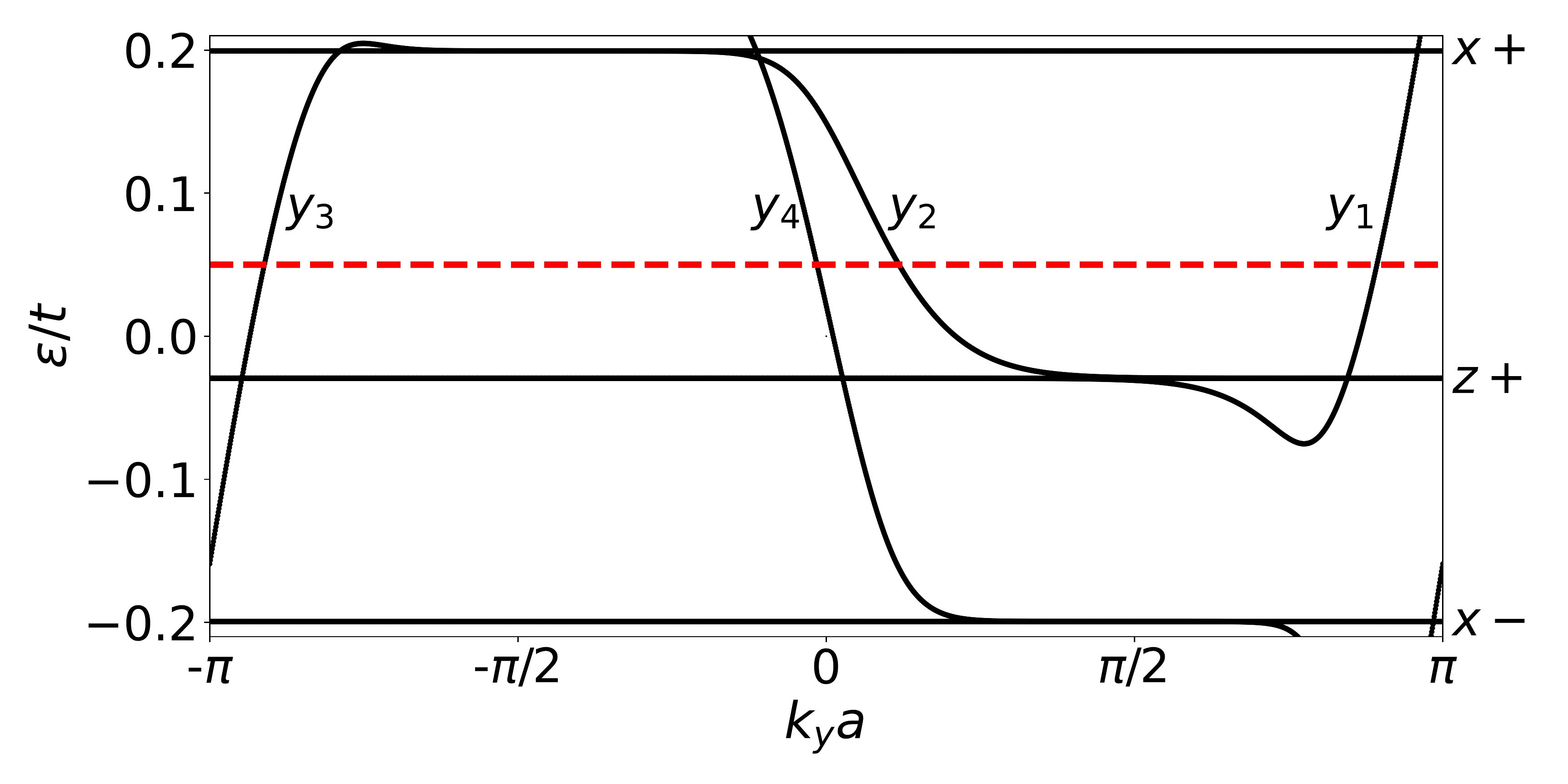}\\
    \includegraphics[width=\columnwidth]{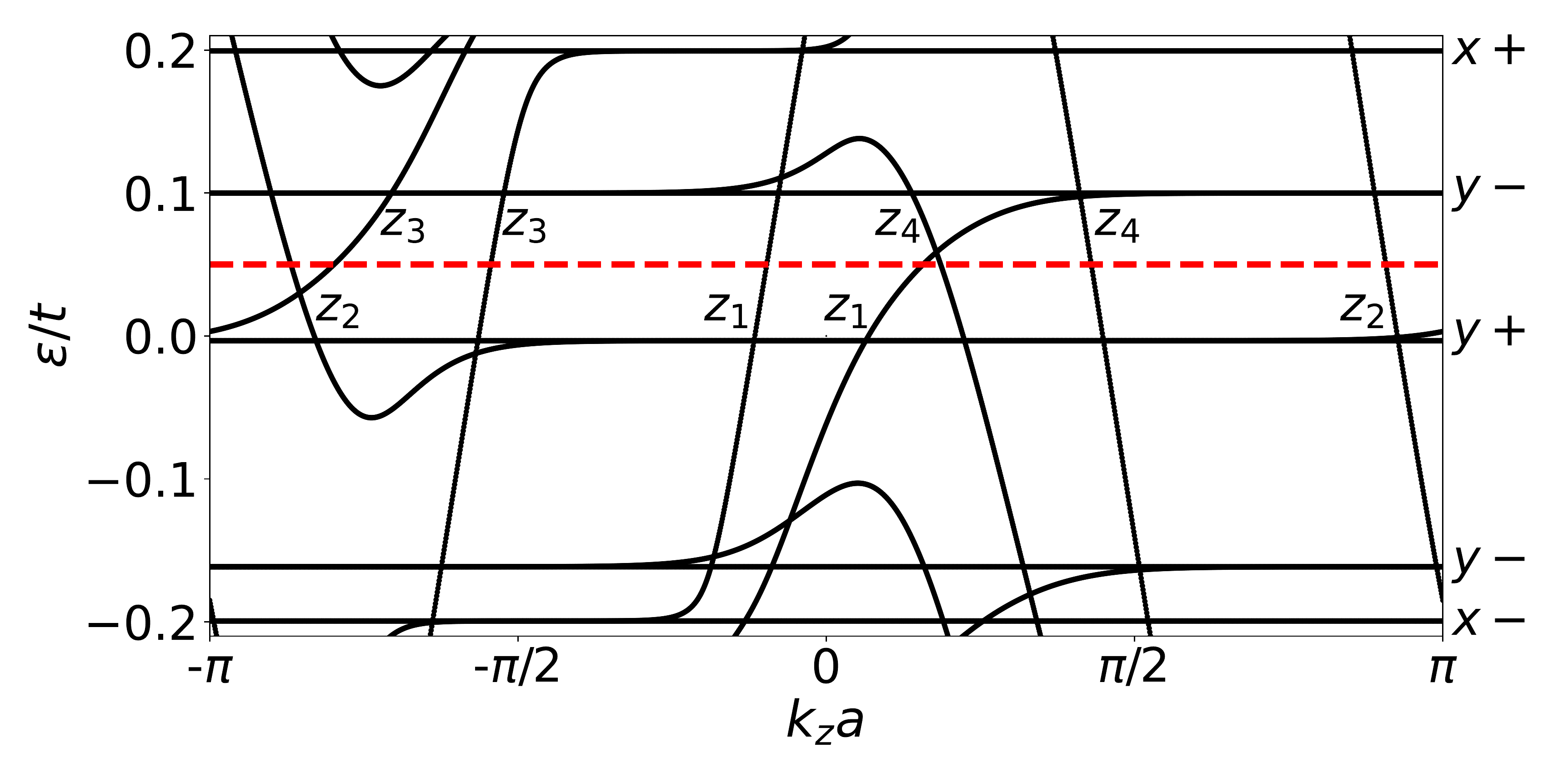}
    \caption{Band structure of a lattice infinite along the $x$ axis (top), $y$ axis (middle), and $z$ axis (bottom) and with a finite size in the other two coordinate directions. The flat bands correspond to the surface Landau levels. The dispersing bands are hinge states. The hinge modes are labelled by which hinge they appear on, and bulk surface modes by their surface, following the convention of Fig.\ \ref{fig:model} (right panel). Parameter values are described in the main text. The lattice sizes in the directions in which they are finite are $L_x = 50a$, $L_y = 80a$, and $L_z = 80a$. The horizontal dashed line indicates the Fermi energy $\varepsilon_{\rm F} = 0.05t$. The zeroth Landau level for the $z-$ surface is at $\varepsilon=0.3t$, which is outside the range shown in the figure.
    \label{fig:bandstructures}}
\end{figure}

To establish that the model describes an extrinsic second-order topological insulator \cite{Sitte2012} with the configuration of hinge states shown in Fig.\ \ref{fig:schematic}, we consider a system that is infinite along each one of the coordinate axes and calculate the corresponding band structure. Examples of such band structures in the vicinity of the Fermi energy are shown in Fig.\ \ref{fig:bandstructures}. In Fig.\ \ref{fig:bandstructures} one easily recognizes the flat surface bands of the Landau levels and the dispersing one-dimensional hinge states. For these numerical calculations, the magnetic field $\vB_0$ was set to $\vB_0 = ({h c}/{60 e a^2}) (2 \sqrt{2} \cos 67^\circ, 2 \sqrt{2} \sin 67^\circ, 3)$ and the Fermi energy was set to $\varepsilon_{\rm F} = 0.05 t$ (indicated by the horizontal dashed line in Fig.\ \ref{fig:bandstructures}). The gate voltages are $V_{x+} = 0.2t $, $V_{x-} = -0.2t $, $V_{y+} = -0.7t $, $V_{y-} = 0.8t $, $V_{z+} = -0.03t $, $V_{z-} = 0.3t $, where we used the convention of Fig.\ \ref{fig:model} (right) to label the surfaces. Whereas the detailed band structures shown in Fig.\ \ref{fig:bandstructures} depend on the gauge choice made for the vector potential $\vA_0(\vr)$ used to describe the uniform magnetic field $\vB_0$, the numbers of hinge modes at each crystal edge, their velocities, and, if applicable, the momentum difference between co-propagating hinge modes at the same hinge do not depend on it. Band structures covering a larger range of energies are shown in App.\ \ref{app:bandstructures}. 

\section{Mach-Zehnder Interferometer}
\label{sec:3}

\begin{figure}
    \centering
    \includegraphics[width=0.95\columnwidth]{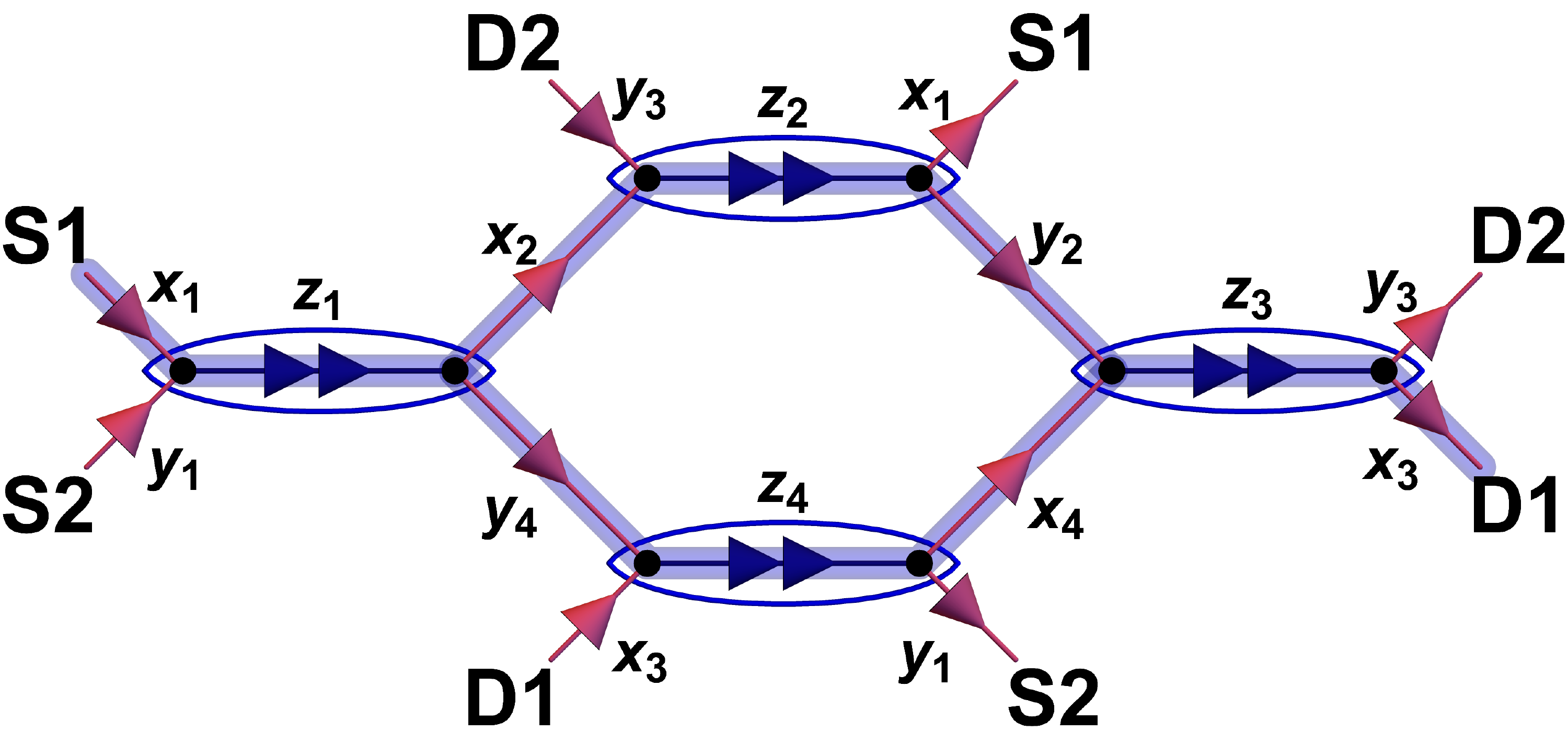}
    \caption{Effective network diagram for the Mach-Zehnder interferometer of Fig.\ \ref{fig:schematic} with labeling of the crystal edges following the convention of Fig.\ \ref{fig:model}. The interfering paths from source contact S1 to drain contact D1 are indicated in blue. The edges with two co-propagating hinge modes serve as the beam splitters, as indicated by the blue ovals.}
    \label{fig:network}
\end{figure}

To construct a Mach-Zehnder interferometer, Ohmic contacts S1, S2, D1, and D2 are added to four of the crystal edges that support only a single chiral hinge mode, see Fig.\ \ref{fig:schematic}. Using the labeling convention of Fig.\ \ref{fig:model} (right), these are the edges $x1$, $x3$, $y1$, and $y3$. The effective network diagram of Fig.\ \ref{fig:schematic} (right) is reproduced in Fig.\ \ref{fig:network}, with the labels of the individual edges added. 

A bias voltage $V$ is applied to the source contact S1, whereas the other three contacts are kept grounded. The current $I$ in response to the bias voltage is measured in drain contact D1. Using the Landauer-B\"uttiker formalism \cite{Buttiker1986}, the conductance $G = I/V$ may be expressed in terms of the $4 \times 4$ scattering matrix $S$ of the system,
\begin{equation}
  G = \frac{e^2}{h} |S_{D1,S1}|^2.
  \label{eq:G}
\end{equation}
We note that $S$ is a $4 \times 4$ matrix even for Ohmic contacts, which have many channels, because the number of channels coupling to the system is limited by the number of chiral modes at the hinge connected to the Ohmic contacts.

The scattering matrix $S$ may be expressed in terms of $2 \times 2$ scattering matrices of the eight crystal corners and in terms of scattering phases accumulated along the crystal edges. Hereto we first construct $2 \times 2$ scattering matrices of the four edges $z\alpha$ with two co-propagating hinge modes, $\alpha=1,2,3,4$, which serve as beam splitters in the interferometer network, see Fig.\ \ref{fig:network}. (We refer to Fig.\ \ref{fig:model} (right) for  the labeling convention for the hinges.) The beam-splitter scattering matrices are denoted $t_{z\alpha}$, $\alpha=1,2,3,4$. Each of these is the product of $2 \times 2$ scattering matrices $t_{z\alpha}^+$ and $t_{z\alpha}^-$ of the crystal corners of the crystal edge $z\alpha$ and a diagonal matrix containing the phases $\phi_{z\alpha,1}$ and $\phi_{z\alpha,2}$ accumulated by the two chiral modes at the edge $z\alpha$,
\begin{align}
  \label{eq:tz}
  t_{z\alpha} = &\, \left\{ \begin{array}{ll}
    t_{z\alpha}^- \mbox{diag}\, (e^{i\phi_{z\alpha,1}},e^{i\phi_{z\alpha,2}}) t_{z\alpha}^+ & \alpha = 1,3, \\
    t_{z\alpha}^+ \mbox{diag}\, (e^{i\phi_{z\alpha,1}},e^{i\phi_{z\alpha,2}}) t_{z\alpha}^- & \alpha = 2,4. \end{array} \right.
\end{align}
Arranging the rows and columns of the $2 \times 2$ matrices $t_{z\alpha}$ such that the first (second) row/column corresponds to an outgoing/incoming state at an edge parallel to the $x$ ($y$) axis, we then find
\begin{align}
  \label{eq:S}
  |S_{D1,S1}|^2 =& |t_{z3,12} e^{i \phi_{y2}} t_{z2,21} e^{i \phi_{x2}} t_{z1,11} \nonumber \\ &\, \ \ \mbox{} +
  t_{z3,11} e^{i \phi_{x4}} t_{z4,12} e^{i \phi_{y4}} t_{z1,21}|^2.
\end{align}

We write the propagation phases $\phi_{z\alpha,1}$ and $\phi_{z\alpha,2}$ for propagation along the edges with two co-propagating hinge modes as
\begin{align}
  \label{eq:dz}
  \phi_{z\alpha,1} =&\, \phi_{z\alpha} + \frac{1}{2} \Delta k_{z\alpha} L_z,\nonumber \\
  \phi_{z\alpha,2} =&\, \phi_{z\alpha} - \frac{1}{2} \Delta k_{z\alpha} L_z,
\end{align}
where $L_z$ is the crystal size in the $z$ direction ({\em i.e.}, along the crystal edges with two co-propagating hinge modes) and $\Delta k_{z\alpha}$ is the momentum difference between the two co-propagating hinge modes at the edge $z\alpha$. The momentum difference $\Delta k_{z\alpha}$ is gauge independent and can be obtained from the one-dimensional band structures shown in Fig.\ \ref{fig:bandstructures}. It does not change under the small changes $\delta \vB=(B_x,B_y,B_z)$ of the applied magnetic field required to observe the interference pattern, because this field scale is proportional to the sample cross section, whereas the field dependence of $\Delta k_{z\alpha}$ is on a scale proportional to the sample length $L_z$.
The phases $\phi_{z\alpha}$, $\alpha=1,2,3,4$, $\phi_{x2}$, $\phi_{x4}$, $\phi_{y2}$, and $\phi_{y4}$ are gauge dependent. However, the conductance $G$ depends on the combination $\phi = \phi_{x2} + \phi_{z2} + \phi_{y2} - \phi_{x4} - \phi_{z4} - \phi_{y4}$ only, which is gauge independent and depends linearly on the total magnetic flux $\Phi$ enclosed between the two interfering paths,
\begin{equation}
  \phi = 2 \pi \Phi/\Phi_0 + \mbox{const}.,
  \label{eq:phi}
\end{equation}
where $\Phi_0 = 2 \pi \hbar c/e$ is the flux quantum. For the geometry of Fig.\ \ref{fig:schematic}, one has 
\begin{equation}
  \Phi = B_z L_x L_y+B_x L_y L_z+B_y L_z L_x,
  \label{eq:Fluxdef}
\end{equation}
where $L_x$ and $L_y$ are the system dimensions in the $x$ and $y$ directions, respectively. Substituting Eqs.\ \eqref{eq:tz}--\eqref{eq:phi} into Eq.\ \eqref{eq:G} one obtains the sinusoidal magnetic-field dependence of the conductance characteristic of a Mach-Zehnder interferometer.

\begin{figure}
    \centering
    \includegraphics[width=\columnwidth]{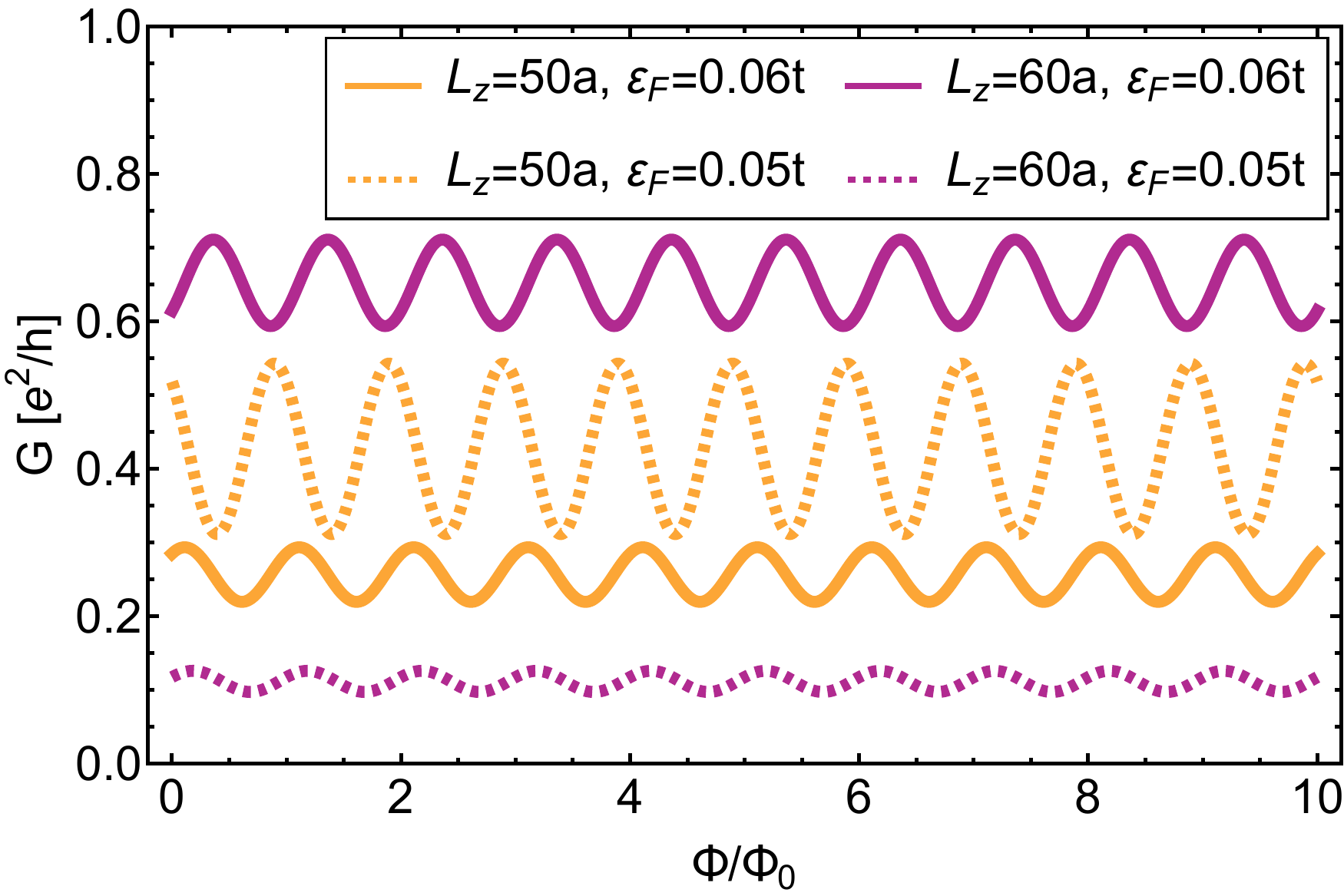}
    \caption{Differential conductance $G$ versus magnetic flux $\Phi = B_z L_x L_y$ for the Mach-Zehnder geometry of Fig.\ \ref{fig:schematic} (left) with $B_y=B_x=0$. Curves are shown for two different values of the Fermi energy $\varepsilon_{\rm F}$ and two different values of the system size $L_z$, as indicated in the figure.}
    \label{fig:cond}
\end{figure}

In App.\ \ref{app:kwant} we describe how the eight scattering matrices $t_{z\alpha}^{\pm}$ of individual corners can be calculated for the lattice model of Sec.\ \ref{sec:2} using the kwant software \cite{Groth2014}, up to two magnetic-field independent over-all phase factors that can be absorbed into the propagation phases $\phi_{x\beta}$ and $\phi_{y\beta}$ of the crystal edges with single chiral modes. With this knowledge, the full interference pattern for the conductance $G$ can be calculated for the model of Sec.\ \ref{sec:2}. Examples of interference patterns for different Fermi energies and values of $L_z$ are shown in Fig.\ \ref{fig:cond}.

In a Mach-Zehnder interferometer, the interference contrast is determined by the properties of the beam splitter. For the interferometer considered here, the scattering matrices $t_{z\alpha}$ of the ``beam splitters'' can be manipulated externally via the momentum difference $\Delta k_{z\alpha}$, see Eqs.\ \eqref{eq:tz} and \eqref{eq:dz}. Small variations of $\Delta k_{z\alpha}$ may have a large effect on $t_{z\alpha}$ because of the presence of the macroscopic factors $L_{z}$ in Eq.\ \eqref{eq:dz}. The momentum difference $\Delta k_{z\alpha}$ depends on the Fermi energy $\varepsilon_{\rm F}$ and on the gate voltages applied to the adjacent surfaces. Indeed, Fig.\ \ref{fig:cond} shows different interference patterns for interferometers with different Fermi energy $\varepsilon_{\rm F}$. Adding disorder to the system will change the particular values of the beam-spitter scattering matrices $t_{z\alpha}$, but will not systematically affect the results or suppress the interference contrast, provided the bulk gaps remain open.

\section{Two-terminal Aharonov-Bohm interferometer}
\label{sec:4}

It is the presence of four Ohmic contacts in the geometry of Fig.\ \ref{fig:schematic} that limits the number of interfering paths between a given pair of source and drain contacts to two and, hence, leads to the characteristic sinusoidal interference pattern characteristic of a Mach-Zehnder interferometer. A two-terminal geometry, with only a single source and a single drain contact, allows for multiple interference paths and, hence, has a more complicated interference pattern. In this Section, we discuss results obtained for such a two-terminal interferometer.

\begin{figure}
  \centering
\includegraphics[width=0.45\columnwidth]{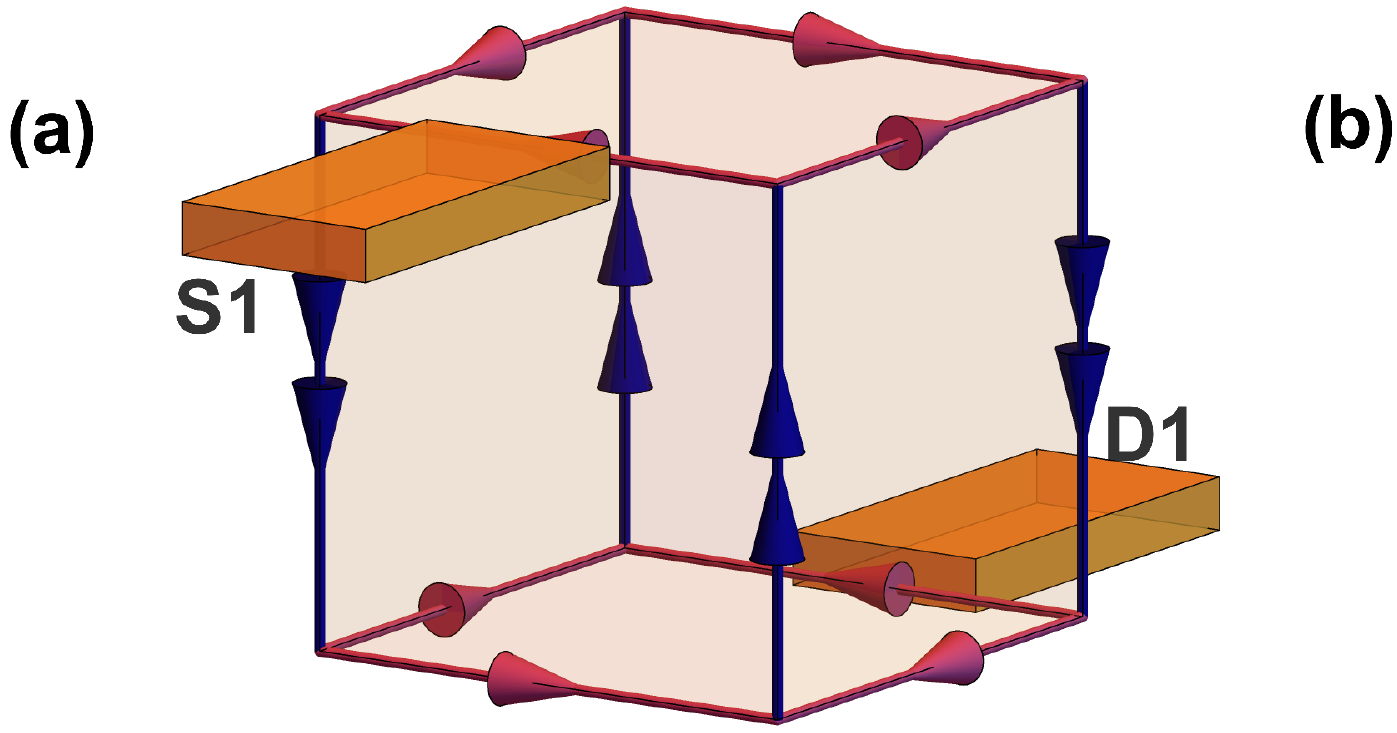}
\includegraphics[width=0.45\columnwidth]{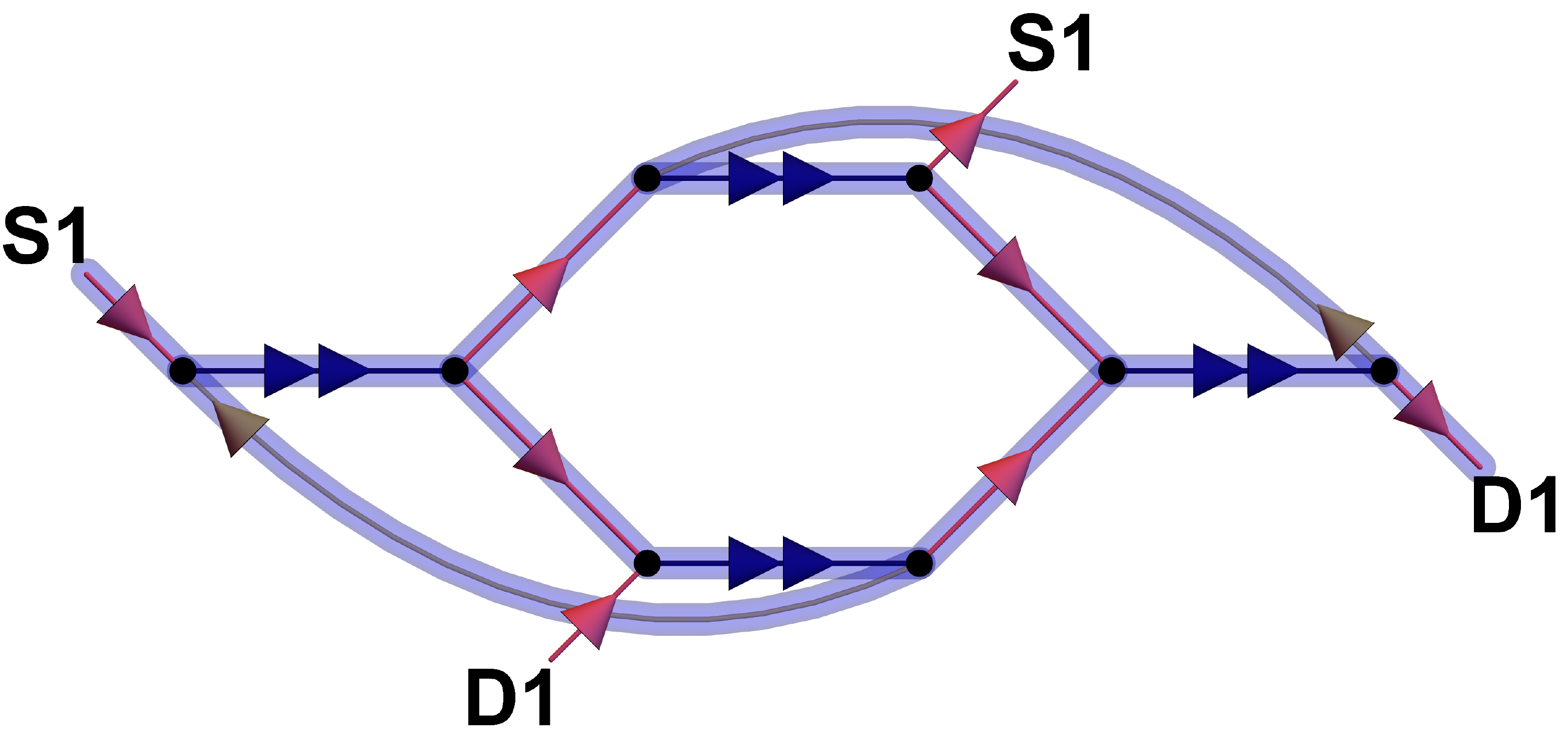}
    \caption{Two-terminal Aharonov-Bohm interferometer constructed from a second-order topological insulator with hinges that have one or two hinge states (left) and equivalent network diagram (right).}
    \label{fig:schematic2}
\end{figure}

A schematic of the two-terminal interferometer is shown in Fig.\ \ref{fig:schematic2}, together with an effective network diagram. In the two-terminal geometry, the conductance $G$ is still given by Eq.\ \eqref{eq:G}, but $S$ now is a $2 \times 2$ matrix and its calculation in terms of the scattering matrices $t_{z\alpha}^{\pm}$ of the crystal corners and the phases accumulated along the hinges without Ohmic contacts is more involved than in the four-terminal case considered in Sec.\ \ref{sec:3}.

In this section we calculate the source-drain current in the two-terminal interferometer geometry. Hereto, we first define the auxiliary $2 \times 2$ matrices
\begin{align}
  (\tau^{\rm u})_{ml} =&\, (t^+_{z3})_{m2} e^{i \phi_{y2}} (t_{z2})_{21} e^{i \phi_{x2}} (t^-_{z1})_{1l}, \nonumber \\
  (\tau^{\rm l})_{ml} =&\, (t^+_{z3})_{m1} e^{i \phi_{x4}} (t_{z4})_{12} e^{i \phi_{y4}} (t^-_{z1})_{2l}, 
\end{align}
which contain the propagation amplitudes for the two paths linking the crystal edges $z1$ and $z3$ with each other. With this notation, the scattering matrix element $S_{D1,S1}$ describing the Mach-Zehnder interferometer with four contacts has the simple form (compare with Eq.\ \eqref{eq:S})
\begin{equation}
  S_{D1,S1} = [t_{z3}^- (\tau^{\rm u} + \tau^{\rm l}) t_{z1}^+]_{11}.
\end{equation}
In addition to the paths linking the edges $z1$ and $z3$ with each other, in the two-terminal geometry also loops linking the edges $z1$ and $z3$ to themselves are possible, see Fig.\ \ref{fig:schematic2}(b). These are described by the $2 \times 2$ matrices
\begin{align}
  (\rho^{\rm ll})_{ml} =&\, (t^+_{z1})_{m2} e^{i \phi_{y1}} (t_{z4})_{22} e^{i \phi_{y4}} (t^-_{z1})_{2l}, \nonumber \\
  (\rho^{\rm uu})_{ml} =&\, (t^+_{z3})_{m2} e^{i \phi_{y2}} (t_{z4})_{22} e^{i \phi_{y3}} (t^-_{z3})_{2l},
\end{align}
which describe loops around the faces $x+$ and $x-$, respectively. With this notation, the scattering matrix element $S_{D1,S1}$ describing the Mach-Zehnder interferometer with two contacts reads
\begin{align}
  S_{D1,S1} = \left[ 
  t_{z3}^- (1 - \rho^{\rm uu})^{-1} (\tau^{\rm u} + \tau^{\rm l}) 
  (1 - \rho^{\rm ll})^{-1} t_{z1}^+
  \right]_{11}.
\end{align}

In order to bring about the interference patterns, it is convenient to keep the accumulated phases from the magnetic field explicit. We define $\phi_x = 2 \pi \delta B_x L_y L_z/\Phi_0$, $\phi_{y} = 2 \pi \delta B_y L_z L_x/\Phi_0$, and $\phi_z = 2 \pi \delta B_z L_x L_y/\Phi_0$, where $\delta \vB$ is the small shift of the magnetic field used to obtain the interference pattern, and obtain
\begin{align}
  |S_{D1,S1}|^2 = \left| \sum_{\alpha = -\infty}^{\infty}
   e^{i (1+\alpha) \phi_x + i \phi_y + i \phi_z} A_{\alpha}
   + A_{\alpha}' 
  \right|^2,
  \label{eq:SAexpand}
\end{align}
with
\begin{align}
  A_{\alpha} =&\, \sum_{\beta} [t_{z3}^- 
  (\rho^{\rm uu})^{\beta+\alpha} \tau^{\rm u}
  (\rho^{\rm ll})^{\beta} t_{z1}^+]_{11}, \nonumber \\
  A_{\alpha}' =&\,  \sum_{\beta} [t_{z3}^- 
  (\rho^{\rm uu})^{\beta} \tau^{\rm l}
  (\rho^{\rm ll})^{\beta + \alpha} t_{z1}^+]_{11},
\end{align}
where the summation is restricted to those values of $\beta$ for which $\beta$ and $\beta + \alpha$ are non-negative and the scattering matrices are evaluated for $\delta \vB = 0$. Eq.\ \eqref{eq:SAexpand} clearly shows the possible phases contributing to the Aharonov-Bohm effect. These phases can be directly read off in the Fourier transforms of the interference patterns.

\begin{figure*}
    \centering
 \includegraphics[width=2\columnwidth]{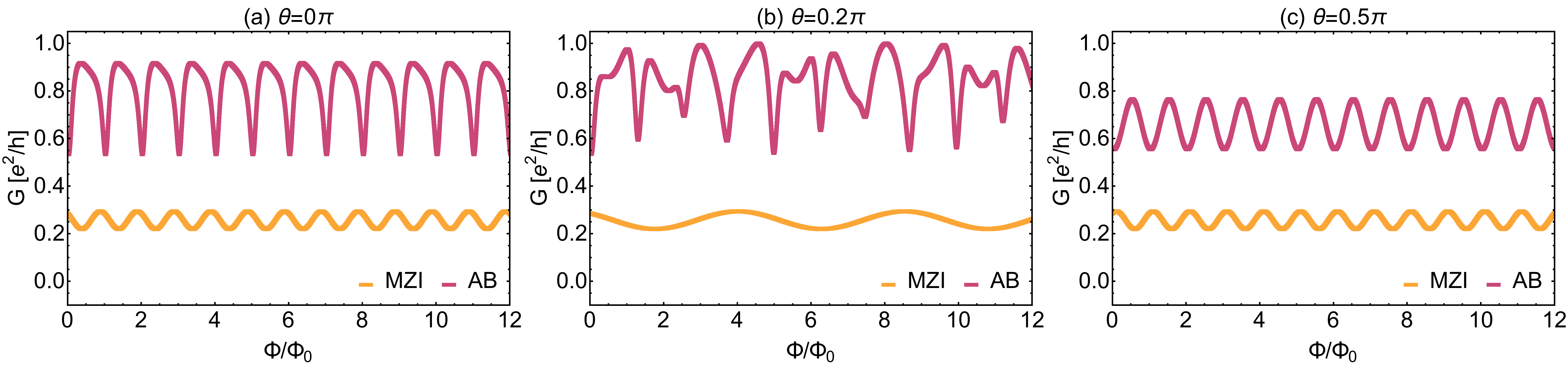}
\caption{Differential conductance $G$ versus magnetic flux $\Phi$ for the two-terminal geometry of Fig.\ \ref{fig:schematic2} (red curves) and for the four-terminal geometry of Fig.\ \ref{fig:schematic} (orange curves). The three panels correspond to three directions of the magnetic field variation: $\delta \vB \propto (\cos \theta,0,\sin \theta)$, with $\theta = 0$ (left), $\theta = 0.2 \pi$, (center), and $\theta = \pi/2$ (right). System parameters are $L_x=L_y=L_z=50a$ and $\varepsilon_{\rm F}=0.06t$. The magnetic flux $\Phi$ is the sum of the fluxes through the three crystal faces, see Eq.\ \eqref{eq:Fluxdef}.}
    \label{fig:condarb}
\end{figure*}

In contrast to the Mach-Zehnder interferometer of Fig.\ \ref{fig:schematic}, for which the interference pattern depends on a single component of the magnetic field variation $\delta \vB$ only, the interference pattern of the two-terminal interferometer involves the full vector $\delta \vB$. In Fig.\ \ref{fig:condarb} we show exemplary data for the conductance $G$ vs.\ $\delta \vB$ for a magnetic field variation of the form $\delta \vB \propto (\cos \theta,0,\sin \theta)$, for different angles $\theta$. For $\delta \vB$ along a coordinate axis, {\em i.e.}, $\theta = 0$ or $\theta = \pi/2$, the interference pattern of the two-terminal interferometer is periodic, but not sinusoidal. For generic $\theta$ the interference pattern is generically aperiodic, because periods corresponding to different interference loops are incommensurate.

\begin{figure}
    \centering
    \includegraphics[width=\columnwidth]{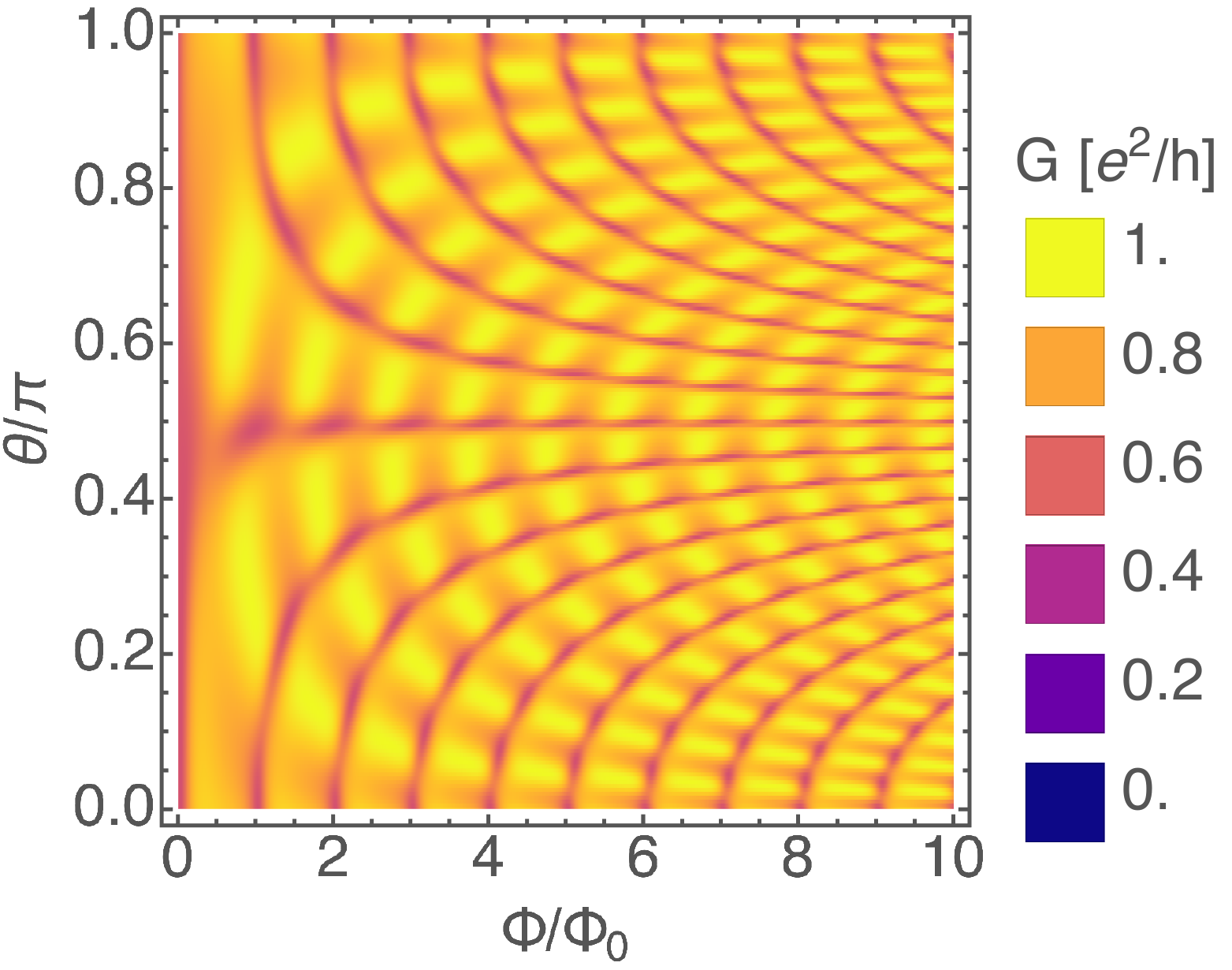}
    \caption{Interference pattern of the differential conductance for an additional field $\delta\vB\propto(\cos\theta,0,\sin\theta)$. System parameters are $L_x=L_y=L_z=50a$ and $\varepsilon_{\rm F}=0.06t$.}
    \label{fig:app0}
\end{figure}

For the Fourier transforms, when the magnetic field $\delta\vB$ is perpendicular to the $x$ axis, one has $\phi_x = 0$ and there is only one peak. On the other hand, for a magnetic field along $x$, Eq.\ \eqref{eq:SAexpand} predicts a flux dependence with many harmonics. In Fig.\ \ref{fig:condarb} of the main text and in Fig.\ \ref{fig:app0} we show examples of the dependence of the two-terminal conductance on the direction of the additional field $\delta\vB\propto(\cos\theta,0,\sin\theta)$ in the $xz$ plane and on its strength (which is parameterized by the total flux $\Phi$ through the three crystal faces, see Eq.\ \eqref{eq:Fluxdef}).

\begin{figure*}
    \centering
    \includegraphics[width=2\columnwidth]{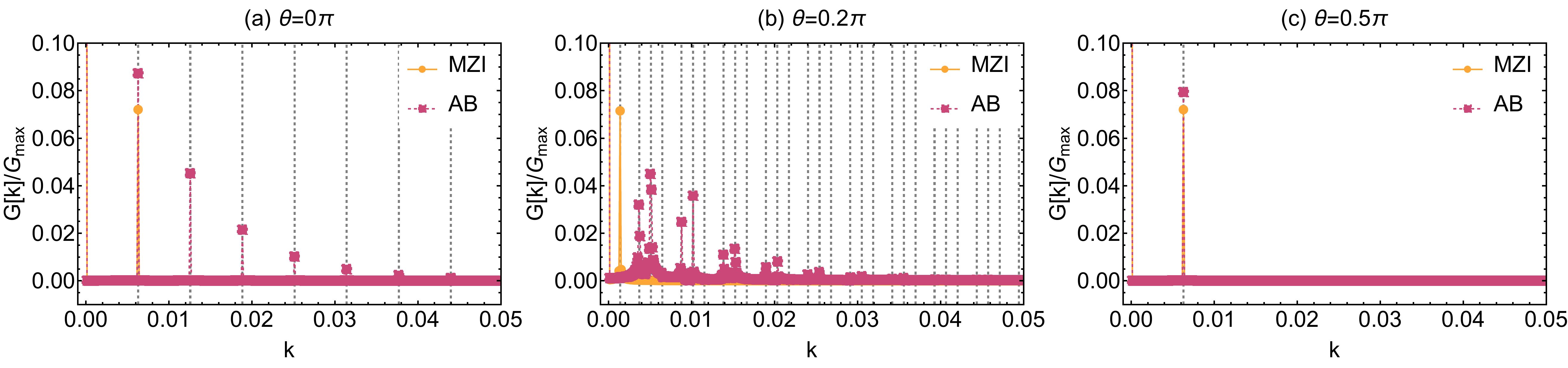}
    \caption{Fourier transform of the differential conductance for an additional field $\delta\vB\propto(\cos\theta,0,\sin\theta)$. Results for the four-terminal (MZI, yellow) and two-terminal (AB, red) cases are shown. Dashed vertical black lines show the locations of the exact peaks in the Fourier transform. Parameter values are as in Fig.\ \ref{fig:condarb}.}
    \label{fig:condarbk}
\end{figure*}

The enclosed areas can be revealed by Fourier transform to the flux $\Phi$. This is illustrated in Fig.\ \ref{fig:condarbk}, where we show the Fourier transform of the two-terminal and four-terminal examples used in Fig.\ \ref{fig:condarb}.

\section{Conclusions and discussion}
\label{sec:5}

In this article we have demonstrated how to make both a Mach-Zehnder interferometer and an Aharonov-Bohm interferometer using the chiral hinge states of a three dimensional higher order topological insulator. A distinctive feature of our setup is the presence of a pair of co-propagating modes on some of the hinges, which allows the creation of beam splitters for this purpose. Ohmic contacts along specific hinges are used for the incoming and outgoing modes and, in the case of the Mach-Zehnder interferometer, for ensuring that only a single loop is available for the propagating modes.

We introduced a minimal model of the system, and solved the scattering problem for the incoming and outgoing modes at each vertex. The minimal model consists of a four band cubic system with topologically protected surface states. The application of a magnetic field and gate voltages gap out the surfaces and enable tuning of the surface topology to generate the desired configuration of hinge states. The interference patterns are calculated from the network diagram of the paths through the set-up, with scattering matrices calculated for each corner separately. This allows us to consider arbitrary system sizes, without having to sacrifice the accuracy of our numerical calculations. We performed detailed tests and comparisons of two-corner and composite single corner set-ups numerically to ensure the calculations are fully converged and under control.

The Mach-Zehnder interferometer demonstrates the expected oscillations as a function of the magnetic flux through the sample, and we further checked its dependence on applied magnetic field angle and system size. For the Aharonov-Bohm interferometer the path of the particles through the system allows for many additional loops, giving rise to very distinctive interference patterns as a function of applied magnetic field strength and direction, which serve as an experimental test of the phenomenon.

A difference with recent works~\cite{Li2021,Luo2021} is that the setup considered here features a crystal for which all surfaces are gapped and Ohmic contacts to hinges, whereas Refs.\ \onlinecite{Li2021,Luo2021} feature point contacts to surfaces with ungapped surface states, with chiral hinge modes only running along the four hinges joining these surfaces. Contacting the hinges is essential for the four-terminal Mach-Zehnder geometry we consider here. On the other hand, the interference patterns we observe for the two-terminal Aharonov-Bohm geometry are quite similar to those of Refs.\ \onlinecite{Li2021,Luo2021}. A minor difference in this case is that the setup of Refs.\ \onlinecite{Li2021,Luo2021} is limited to magnetic fields parallel to the contact planes, whereas the present setup has a nontrivial interference pattern as a function of the full three dimensional magnetic field vector.


Our modeling of the interferometer involves an {\em extrinsic} higher-order topological insulator. For an extrinsic higher-order phase, the presence of chiral hinge modes relies solely on the crystal termination. In contrast, for an intrinsic higher-order topological insulator, the presence of hinge modes is imposed by the topology of the bulk band structure. Nevertheless, for intrinsic higher-order topological insulators the bulk band structure only partially fixes the number of hinge modes, so that a certain degree of control of the crystal termination remains necessary if higher-order topological insulators are to be used for interferometry purposes \cite{Sitte2012,Geier2018,Trifunovic2018}. The advantage of the fully extrinsic scheme we employ here (first proposed by Sitte {\em et al}.\ \cite{Schindler2018}) is that the hinge modes originate from the Dirac-cone surface states of a parent first-order topological insulator state and, hence, can be controlled by standard means such as an applied magnetic field and electrostatic gate voltages.

\acknowledgments

This work was supported by the Polish National Agency for Academic Exchange (NAWA) under the grant 2PPN/BEK/2020/1/00338/DEC/2 (NS) and by the Deutsche Forschungsgemeinschaft (DFG, German Research Foundation) - Project Number 277101999 - CRC TR 183 (project A03) (AYC and PWB).

\appendix

\section{Further Band Structures}\label{app:bandstructures}

\begin{figure}
  \centering
    \includegraphics[width=\columnwidth]{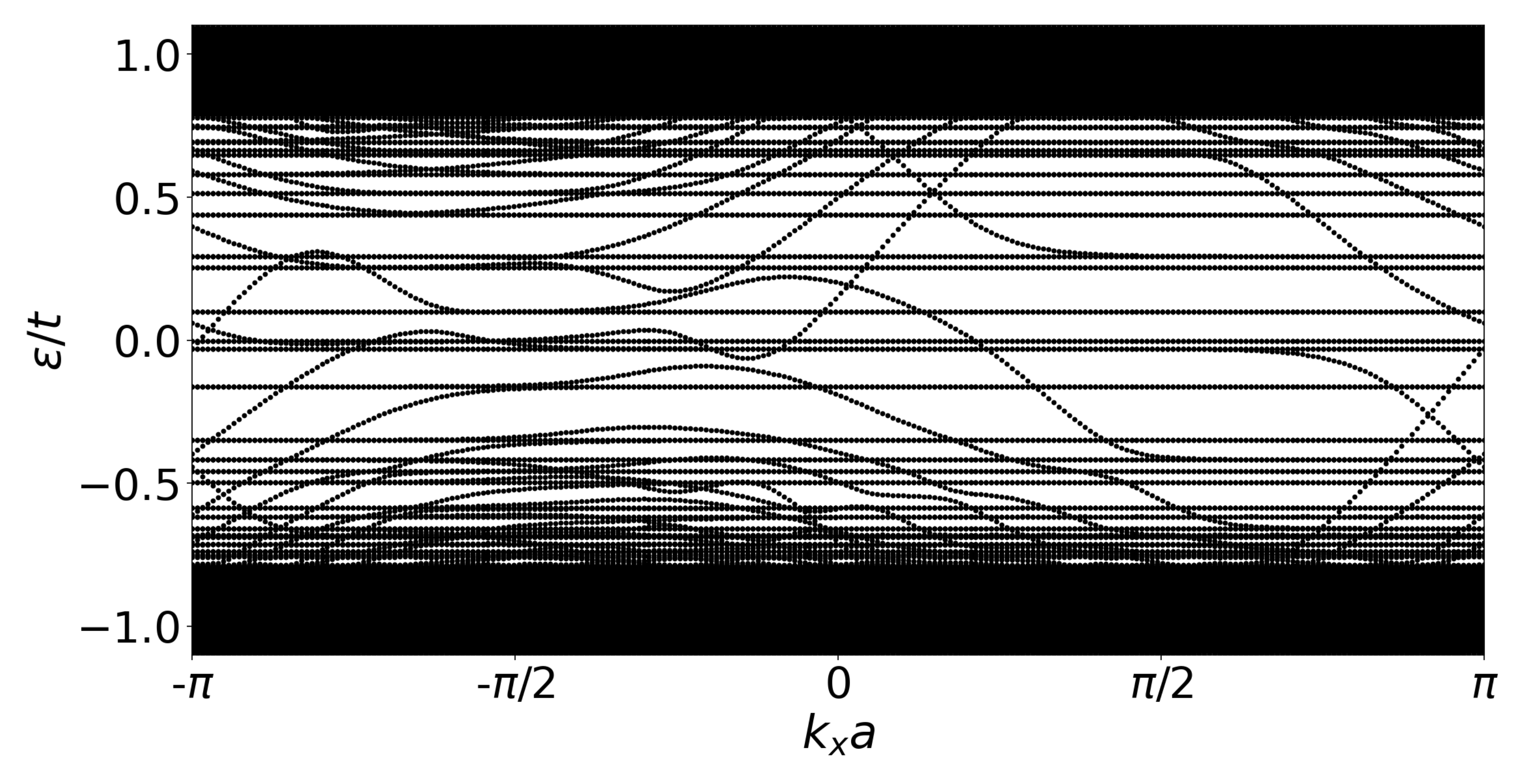}\\
    \includegraphics[width=\columnwidth]{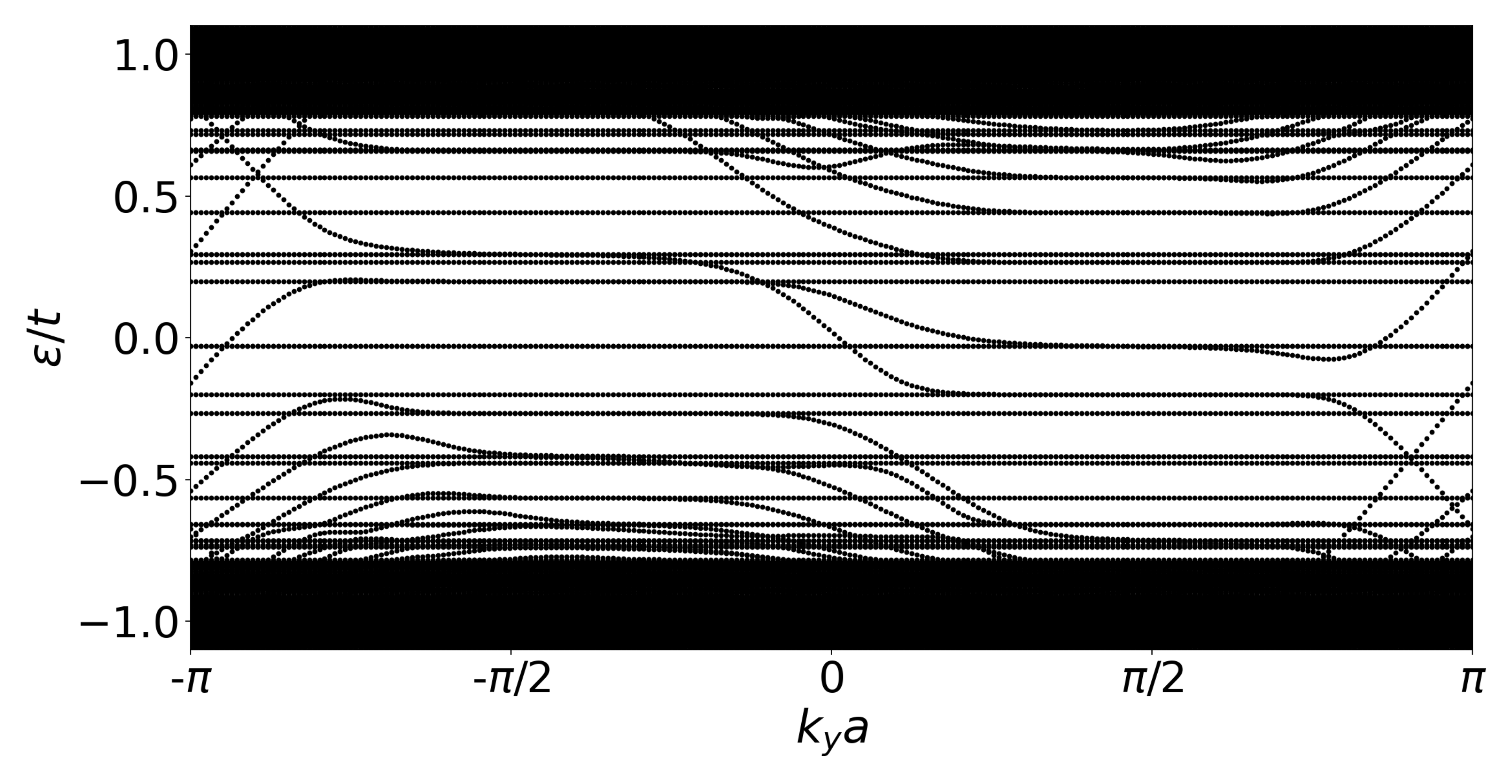}\\
    \includegraphics[width=\columnwidth]{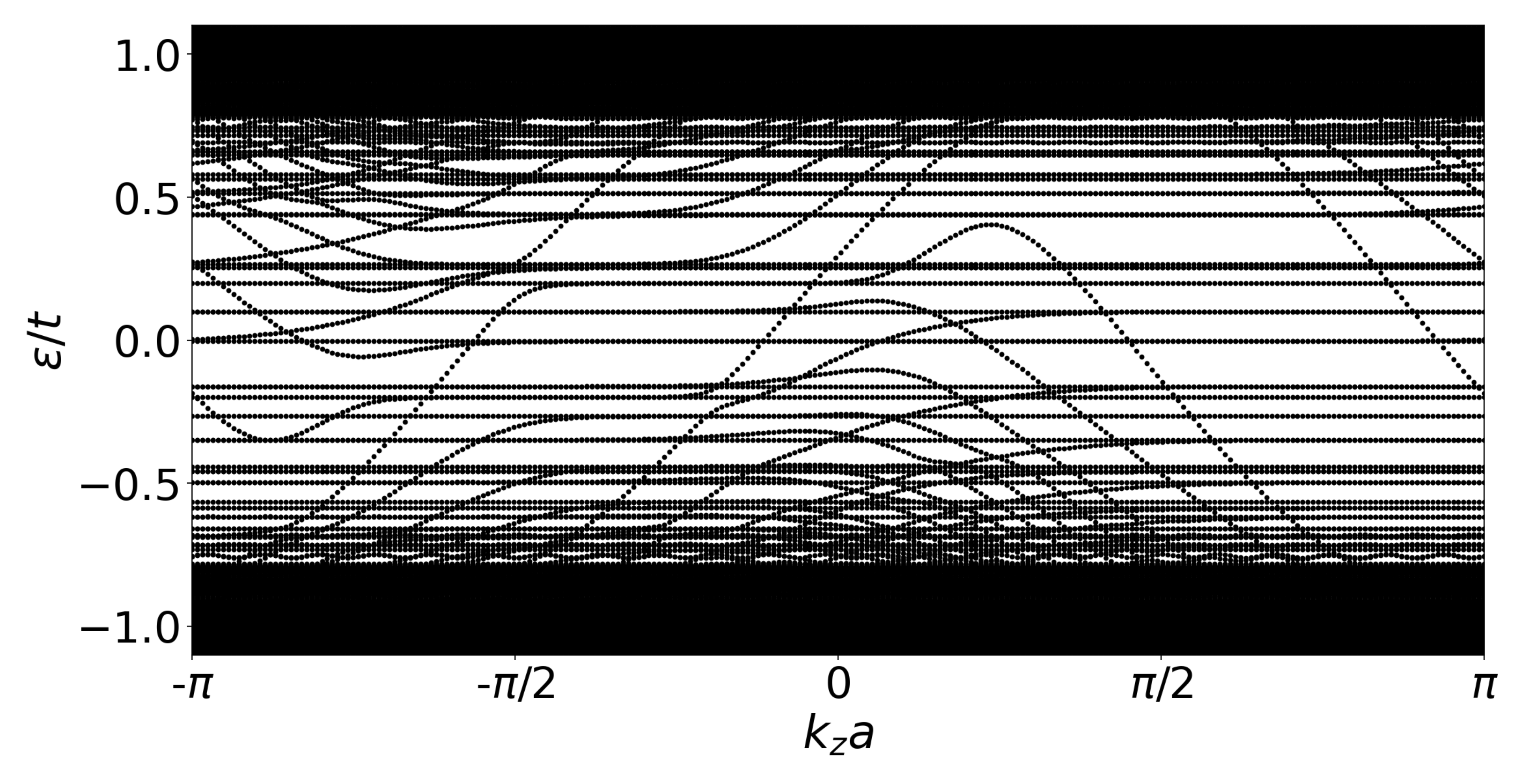}
    \caption{Full band structure in the bulk gap of a lattice infinite along the $x$ axis (top), $y$ axis (middle), and $z$ axis (bottom) and finite along the other two coordinate directions. The dispersing bands are hinge states. The flat bands correspond to the surface Landau levels. The bulk states are visible at the top and bottom of the plots. The parameter values are the same as in Fig.\ \ref{fig:bandstructures}.
    \label{fig:fullgapdispersion}}
\end{figure}

Fig.\ \ref{fig:fullgapdispersion} shows the band structures for a system infinite along the $x$, $y$, or $z$ direction for a larger energy range than shown in Fig.\ \ref{fig:bandstructures} of the main text. In this wider energy range, bulk states (the solid blocks at the top and bottom of the plots), surface landau levels (flat bands) and hinge modes (dispersing modes) are all visible.

\section{Corner Scattering Matrices}
\label{app:kwant}

In order to describe arbitrarily large system sizes we characterise corners by the eight scattering matrices $t_{z\alpha}^{\pm}$, $\alpha=1,2,3,4$, describing the scattering between hinge modes at a single corner of the crystal, see Sec.\ \ref{sec:3} of the main text. We can numerically obtain each of these scattering matrices for a lattice model by applying the kwant software \cite{Groth2014} to a geometry in which there is only one corner with a nontrivial scattering matrix. 

\begin{figure}
    \centering
    \includegraphics[width=0.49\columnwidth]{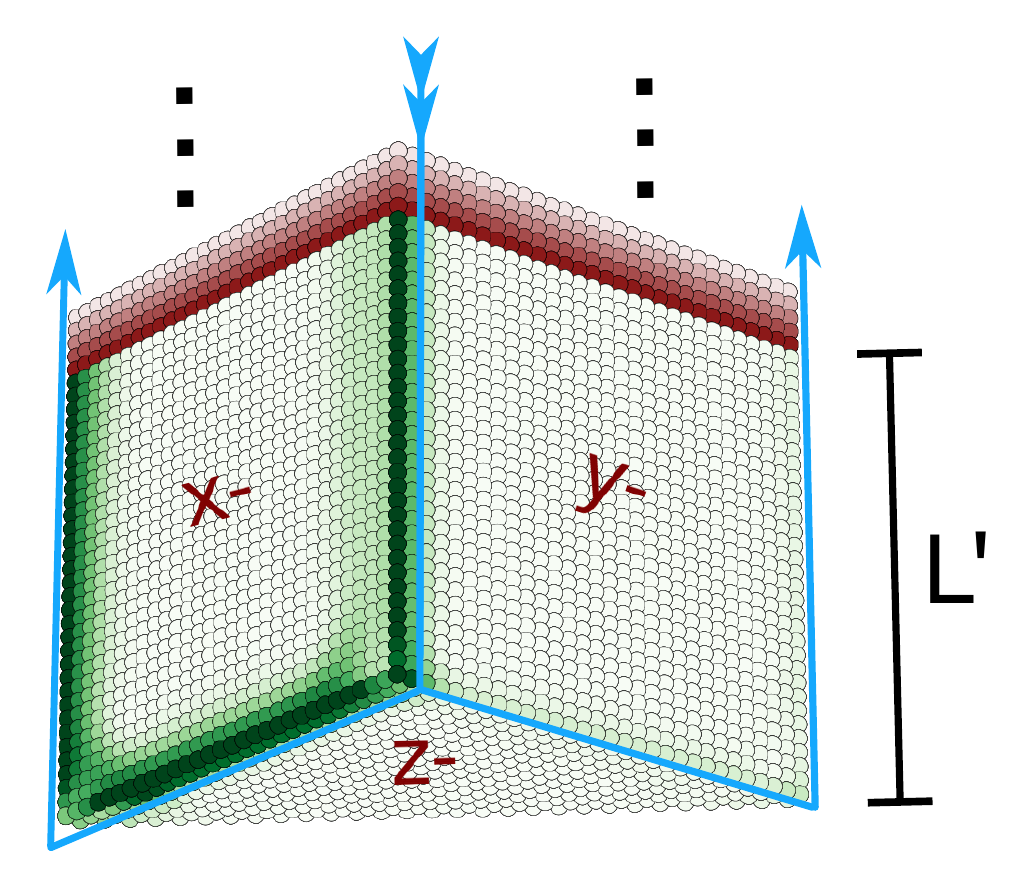}
    \includegraphics[width=0.49\columnwidth]{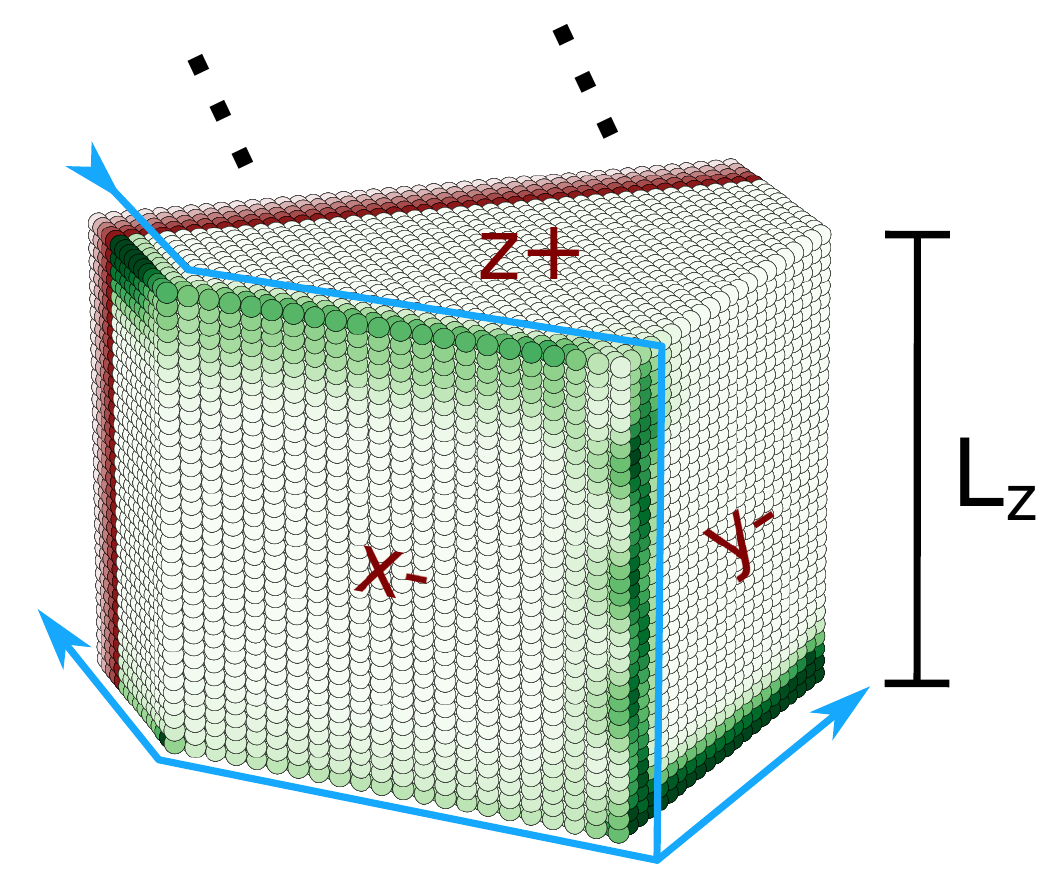}
    \caption{Geometry used to numerically obtain the single-corner scattering matrix $t_{z1}^{-}$ (left panel) and the beam-splitter scattering matrix $t_{z1}$ (right panel) using the kwant software. The red lattice sites represent the ``ideal lead'', which is semi-infinite in one direction (only five layers shown). The color shows the support of a scattering state in the scattering region, at $\varepsilon = 0.077t$. The blue arrows indicate the direction of propagation of the modes. The color scale has been saturated to make both of the outgoing hinge modes more visible. The interference pattern between the two co-propagating modes is clearly visible along the middle hinge in the bottom panel.}
    \label{fig:comp2}
\end{figure}

We illustrate this procedure for the calculation of the scattering matrix $t_{z1}^-$, which describes scattering at the lower left corner of the crystal shown in Fig.\ \ref{fig:schematic}, which is the corner between crystal faces $x-$, $y-$, and $z-$. Figure \ref{fig:comp2} shows the geometry used to calculate this scattering matrix. It consists of a pillar with a triangular cross-section, which is semi-infinite in the $z$ direction. Three faces of the triangular pillar correspond to the crystal faces $x-$, $y-$, and $z-$, whereas the fourth, diagonal face (the back face of the pillar shown in Fig.\ \ref{fig:comp2}) has a termination not present in the crystal of Fig.\ \ref{fig:schematic}. The center hinge, which connects the faces $x-$ and $y-$, supports a pair of modes, while the other two hinges in the $z$ direction and the hinges between the $x-$ and $z-$, as well as $y-$ and $z-$ faces support one mode each. The triangular pillar geometry has three corners. Of these, the corner between the faces $x-$, $y-$, and $z-$ (shown centrally in Fig.\ \ref{fig:comp2}) is the corner of interest. The other two corners, which border on the diagonal face, have one incoming and one outgoing hinge mode, so that they only contribute a phase shift to the scattering state. 

The triangular pillar structure has two incoming modes, propagating along the hinge between the $x-$ and $y-$ faces, and two outgoing modes. Hence, it is described by a $2 \times 2$ scattering matrix $s_{z1}^{-}$, which can be calculated using the standard routines of the kwant software. This scattering matrix is of the form
\begin{equation}
  s_{z1}^{-} = 
  \begin{pmatrix} e^{i \theta_{x}} & 0 \\ 0 & e^{i \theta_{y}} \end{pmatrix}
  t_{z1}^{-} \begin{pmatrix} e^{i k_{z1,1} L'} & 0 \\ 0 & e^{i k_{z1,2} L'} \end{pmatrix},
\end{equation}
where $k_{z1,1}$ and $k_{z2,2}$ are the wavenumbers of the two hinge modes at the hinge between $x-$ and $y-$, $L'$ is the size of the scattering region, and $\theta_x$ and $\theta_y$ are phase shifts accumulated for propagation along hinges and corners with a single mode. Since the wavenumbers $k_{z1,1}$ and $k_{z2,2}$ can be obtained from the dispersion of Fig.\ \ref{fig:bandstructures}, knowledge of $s_{z1}^{-}$ yields $t_{z1}^{-}$ up to left multiplication with a diagonal matrix of phase factors. Not knowing this phase information is unproblematic for the calculation of the ``beam-splitter'' transmission matrices $t_{z\alpha}$ of Eq.\ \eqref{eq:tz}.

\begin{figure}
    \centering
    \includegraphics[width=\columnwidth]{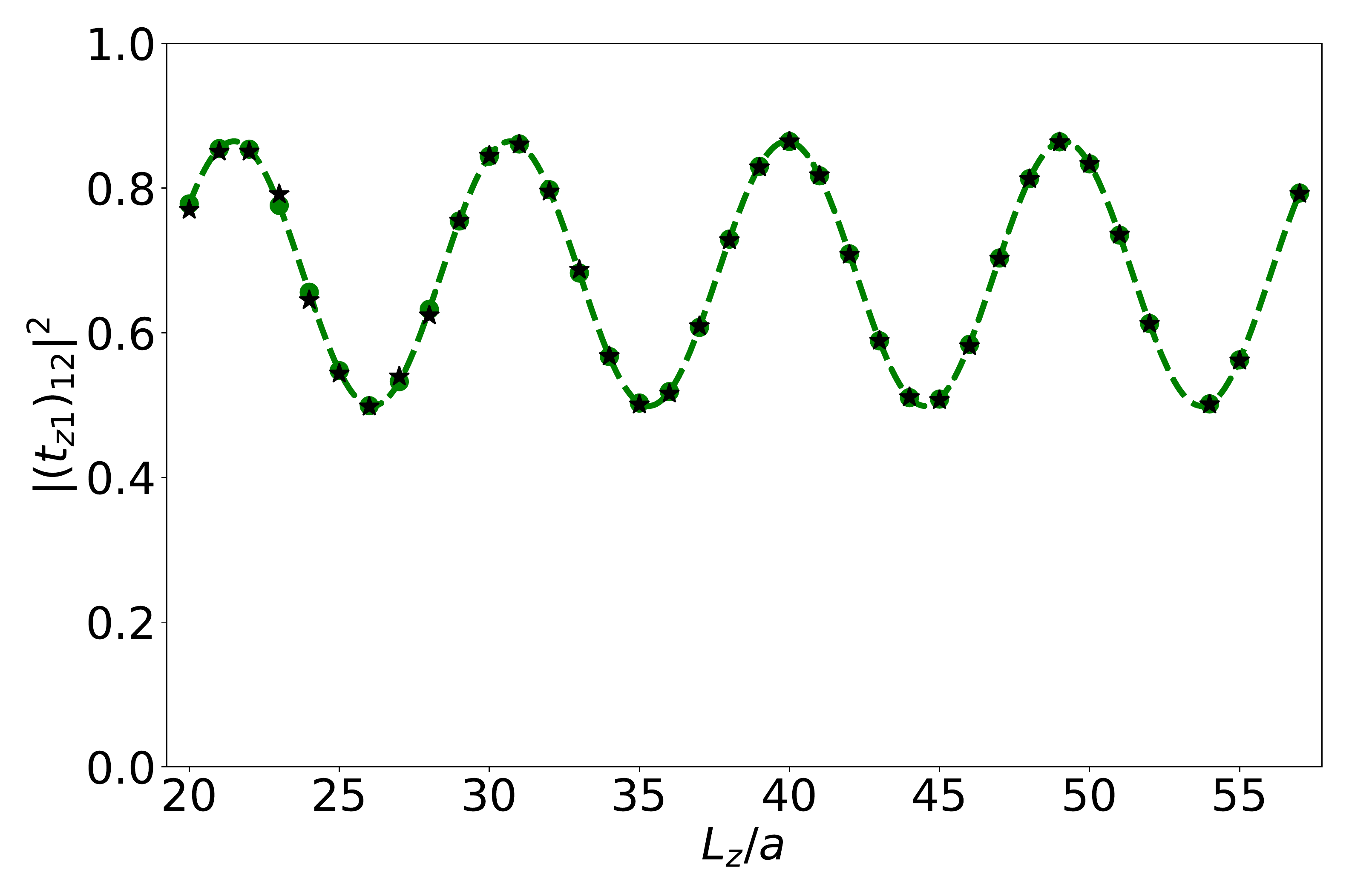}
    \caption{Off-diagonal beam-splitter transmission probability $|(t_{z1})_{12}|^2$ as a function of the separation $L_z$ between corners at at $\varepsilon = 0.077t$. Blue triangle data points are obtained using the geometry of Fig.\ \ref{fig:comp2} bottom panel. Dashed lines and black circles are obtained from Eq.\ \eqref{eq:tz} with single-corner scattering matrices $t_{z1}^{\pm}$ obtained from a geometry as in Fig.\ \ref{fig:comp2} top panel.}
    \label{fig:comp3}
\end{figure}

We have compared building the beam-splitter scattering matrix $t_{z\alpha}$ from the corner scattering matrices $t_{z\alpha}^{+}$ and $t_{z\alpha}^{-}$ and phase shifts for propagation along the crystal hinge in between, see Eq.\ \eqref{eq:tz}, with a direct calculation of $t_{z\alpha}$ using the kwant software. Such a direct calculation is possible by considering a ``wedge-like'' geometry, as shown in Fig.\ \ref{fig:comp2} for the calculation of the beam-splitter scattering matrix $t_{z1}$. Fig.\ \ref{fig:comp3} compares $|(t_{z1})_{12}|^2$ for the two cases and shows good agreement for a separation $\gtrsim 30a$ between the two corners.

%

\end{document}